# What Can Developers' Messages Tell Us?
# A psycholinguistic analysis of Jazz teams' attitudes and behavior patterns

Sherlock A. Licorish and Stephen G. MacDonell

*SERL, School of Computing and Mathematical Sciences*
*Auckland University of Technology*
*Auckland, New Zealand*
*sherlock.licorish@aut.ac.nz, stephen.macdonell@aut.ac.nz*

**Abstract**

*Reports that communication and behavioral issues contribute to inadequately performing software teams have fuelled a wealth of research aimed at understanding the human processes employed during software development. The increasing level of interest in human issues is particularly relevant for agile and global software development approaches that emphasize the importance of people and their interactions during projects. While mature analysis techniques in behavioral psychology have been recommended for studying such issues, particularly when using archives and artifacts, these techniques have rarely been used in software engineering research. We utilize these techniques under an embedded case study approach to examine whether IBM Rational Jazz practitioners' behaviors change over project duration and whether certain tasks affect teams' attitudes and behaviors. We found highest levels of project engagement at project start and completion, as well as increasing levels of team collectiveness as projects progressed. Additionally, Jazz practitioners were most insightful and perceptive at the time of project scoping. Further, Jazz teams' attitudes and behaviors varied in line with the nature of the tasks they were performing. We explain these findings and discuss their implications for software project governance and tool design.*

1. **INTRODUCTION**

Debates over the factors that contribute to or constrain software systems' adequacy and have consequential impacts on project success rates have been longstanding [1]. While many recommendations to adopt various software methodologies and tools have been made [2], questions over the outcomes of software development projects continue to be asked [3]. Previous evidence suggests that people factors manifest themselves in communication and behavioral issues, and these underscore the causes of inadequately performing software teams [4]. Thus, studying these factors and issues should provide fruitful avenues for researchers to better understand the impact of people factors on the software process and to offer recommendations for improvements, which may in turn enhance project performance. In fact, almost irrespective of the reports of inadequately performing teams [5], studying human-related issues would seem to be necessary given the emphasis placed on individuals and interactions [6] and collaboration and coordination [7] by recent software development approaches, and the growing body of research studies dedicated to human interaction, communication and coordination themes [8-9]. Role theories and studies in psychology have shown that various attitudes and behaviors are prevalent and/or necessary in some team environments, while other settings may demand different attitudes for teams to succeed [10]. We are interested in determining whether these findings may also be applicable to software engineering teams [11]. Presuming these requirements are pertinent to software teams, their absence may compromise team performance. Thus, enquiries into software teams' processes could provide definitive and concrete recommendations for how to plan for the staffing of software teams given their development portfolio. In addition to suggestions for software project governance, these understandings would also inform the extension of collaboration and process support tools. We have therefore extended our preliminary study examining the effects of team environment on team behaviors [11]. We extracted and analyzed messages and artifacts associated with the work of ten IBM Rational agile global teams from the Jazz repository to examine if these practitioners' behaviors changed over project duration, and to study the way software teams' attitudes and behaviors varied given the tasks they were undertaking. We found significant differences in the way Jazz teams interacted over different project phases and given their portfolio of tasks.

In the next section we present our theoretical background and survey related works, before stating our specific research questions. We then describe our research methods and measures in Section 3, introducing our techniques and procedures in this section. In Section 4 we present our results and analysis. Section 5 then discusses our findings and outlines the implications of our results, and in Section 6 we consider our study's limitations. Finally, in Section 7 we draw conclusions and outline further research directions.

## 2. THEORETICAL BACKGROUND, RELATED WORK AND RESEARCH QUESTIONS

The compilation of archives recording the textual communication activities of software developers has enabled researchers to study certain aspects of practitioners' social behaviors [12]. For instance, Abreu and Premraj [13] analyzed the Eclipse mailing list and found that increases in communication coincided with a high number of bug-introducing changes, and developers communicated most frequently at release and integration time. Cataldo et al. [14] employed social network analysis (SNA) during their study of a large distributed system and found that those who communicated the most also contributed the most on software tasks. Similarly, Shihab et al. [15] found communication activity to be correlated with software development activity when studying the GNOME project. Nguyen et al. [16] uncovered that 75% of Jazz's core team members actively participated in the project's communication, and distance did not delay communication among team members. Works such as those of Bacchelli et al. [17] and Antoniol et al. [18] have used more complex techniques to analyze email and bug description information. In linking email communications to source code using regular expressions and other information retrieval approaches, Bacchelli et al. [17] found that the least complex approach, considering regular expressions in emails, outperformed more complex probabilistic and vector space models. Antoniol et al. [18] used decision trees, naïve Bayes classifiers and logistic regression to classify bugs based on specific terms used in the descriptions of such tasks, lending encouragement to the use of methods leveraging text analysis.

While such methods and their associated tools have been used previously to understand and predict some aspects of software development [19], only a few studies in this domain have considered examining teams' internal behavioral processes from developers' textual communication. This is in spite of the fact, as noted by Bacchelli et al. [17], that natural language analysis techniques have proved to be effective in generating understanding of software developers' language processes. Apart from our recent preliminary work that uses linguistic analysis techniques to investigate Jazz developers' communication [11], Rigby and Hassan [20] is the only other study that was found to examine aspects of team dynamics using textual communication. In analyzing the communication of the developers involved in the Apache project, Rigby and Hassan [20] uncovered that once the two top developers signaled their intentions to leave the project their communications became more negative and instructive, and they spoke mostly in the future tense and communicated with less positive emotions, when compared to their earlier communications. This study also found variations in communicating behavior after releases. In studying two releases, Rigby and Hassan [20] found that developers' communication was optimistic after the first release, whereas the opposite was evident after the second release. These findings suggest that developers are not committed to such projects once they have decided to leave, and that challenges (such as high incidence of defect reports) and how motivated developers were during development may be responsible for the different feelings displayed after releases. In our preliminary work examining three different projects we also found slight variances in behaviors among those undertaking different forms of software task [11]. Such findings are insightful and support the utility of natural language analysis techniques for understanding human processes, but also point to the need for further large scale exploratory research.

Questions in relation to reliability and validity have also been raised for studies examining open source software (OSS) mailing lists due to the way participants' communications are managed in this environment (i.e., anyone is able to post messages and report bugs to such mailing lists [21]). However, studies such as that of Rigby and Hassan [20] and our own previous work [11] provide useful discovery to encourage systematic application of linguistic analysis techniques to study developers' communication in other controlled environments.

Previous work examining similar datasets comprising Jazz [16] and Microsoft software development artifacts [22] have tended to employ mathematically-based analysis techniques (e.g., SNA and its core measures of density and closeness). These approaches have provided much-needed insights relating to the way software practitioners work. However, there still remain several open questions regarding work in this context – those addressed here consider the way practitioners' behaviors change over project duration and how software teams' attitudes and behaviors vary given the tasks they are undertaking during distributed agile projects.

We believe that it is imperative to examine evidence of the actual interactions and engagements of software practitioners if we are to fully comprehend the true nature of teams working in these sorts of environments. Such work is particularly necessary given the current disposition favoring individuals and interactions as against an emphasis on software variations and changes in artifacts such as requirements documents and code [6]. Evidence in these interactions should shed light on agile and global development processes, should help us to test those initial views that stressed the need for the involvement of highly motivated members for such teams to succeed [23], and should enable us to provide definitive and concrete recommendations on how to plan for the staffing of software teams given their development portfolio.

Evidence from other fields indicates that certain attitudes and behaviors are both prevalent and necessary in some environments or contexts, while other settings demand different capabilities and conditions for teams to succeed. We consider these issues in the context of global software development, at both the project and development phase level. The absence of these specific attitudes or conditions may result in challenges to the success of a software project. Such a position is supported by work on role theories [10] and in human resource management and psychology [24]. However, apart from our preliminary work [11], our search of the literature (covering ACM Digital Library, IEEE Xplore, EI Compendex, Inspec, ScienceDirect and Google Scholar) did not unearth any studies, exploratory or otherwise, considering these issues. We therefore answer the following questions by studying software practitioners' messages:

A. How do globally distributed agile software practitioners' behaviors change over project duration?

B. How does the nature of software tasks affect globally distributed agile teams' attitudes and behaviors?

## 3. METHODS AND MEASURES

We employed an embedded case study design [25] in our analysis of the IBM Rational Jazz Repository. This approach is appropriate for understanding complex human processes by relating them to their context [25], the intent of the work reported here. During our study, mining methods were used for data collection (see subsection B) and the extracted data were then scrutinized using linguistic analysis tools and statistical techniques (see subsection C). In the following subsection (subsection A) we provide a description of the repository used as the data source in our study and then we elaborate on the techniques and procedures employed during our research.

### A. Study Repository

We examined development artifacts from a specific release (1.0.1) of Jazz (based on the IBM[R] Rational[R] Team Concert[TM] (RTC) [1]), a fully functional environment for developing software and for managing the entire software development process [26]. The system includes features for work planning and traceability, software builds, code analysis, bug tracking and version control. Changes to source code in the Jazz environment are permitted only as a consequence of a work item (WI) being created beforehand, such as a bug report, a new feature request or a request to enhance an existing feature (and a history log is maintained for each WI). Team member communication and interaction around WIs are captured by Jazz's comment or message functionality. During development at IBM, project communication, the content explored in this study, was enforced through the use of Jazz itself [16].

The Jazz repository comprised a large amount of process data collected from development and management activities across the USA, Canada and Europe. In Jazz each team has multiple individual roles, with a project leader responsible for the management and coordination of the activities undertaken by the team (team members also work across projects). Jazz teams use the Eclipse-way methodology for guiding the software development process [26]. This methodology outlines iteration cycles that are six to eight weeks in duration, comprising planning, development and stabilizing phases. Builds are executed after project iterations. All information for the software process is stored in a server repository, which is accessible through a web-based or Eclipse-based (RTC) client interface. This consolidated data storage and enforced project controls make the Jazz data more complete and representative of the software process than that in many OSS repositories.

### B. Data Pre-processing and Metrics Definition

Although an investigation of data mining (a broad and vibrant research area in its own right) is beyond the scope of this work, aspects of data mining supported the activities involved in this project in terms of extracting, preparing and exploring the data under observation. Data cleaning, integration and transformation techniques were utilized to maximise the representativeness of the data under consideration and to help with the assurance of data quality, while exploratory data analysis (EDA) techniques were employed to investigate data properties and for anomaly detection. Through these latter activities we were able to identify all records with inconsistent formats and data types, for example: an integer column with an empty cell. We wrote scripts to search for these inconsistent records and tagged those for deletion. This exercise allowed us to identify and delete 122 records (out of 36,672) that were of inconsistent format. We also wrote scripts that removed all HTML tags and foreign characters (as these would have confounded the linguistic analysis).

*1) Data Extraction*

We leveraged the IBM Rational Jazz Client API to extract team information and development and communication artifacts from the Jazz repository. These included (in addition to WIs discussed in subsection A):

1. Project Workspace/Area – each Jazz team is assigned a workspace. The workspace (or project/team area) has all the artifacts belonging to the specific team.

2. Contributors and Teams – a contributor is a practitioner contributing to one or more software features; multiple contributors form teams.

3. Comments or Messages – communication around WIs is captured by Jazz comment functionality. Messages ranged from as short as one word (e.g., thanks) up to 1055 words representing multiple pages of communication.

We extracted the relevant information from the Jazz repository and selected all the artifacts belonging to ten different project areas (out of 94) for analysis. This formed a purposive rather than random sample. Table I shows that the

---
[1] IBM, the IBM logo, ibm.com, and Rational are trademarks or registered trademarks of International Business Machines Corporation in the United States, other countries, or both

project areas selected represented both information-rich and information-rare cases in terms of WIs and messages. Projects ranged from short (59 days) to long (1014 days), with varying levels of communication density. The selected project artifacts amounted to 1201 software development tasks, carried out by a total of 394 contributors working across the ten teams (and comprising 146 distinct members), with 5563 messages exchanged around the 1201 tasks. As the data were analyzed, it became clear that the cases selected were representative of those in the repository, as we reached data saturation [27] after analyzing the third project case. Our use of SNA to initially explore the projects' communication showed a similar graph to that in Fig. 1 for all of the ten project areas (note dense communication segments for specific developers and tasks). Additionally, all ten projects had similar profiles for network density (between 0.02 and 0.14) and closeness (between 0 and 0.06).

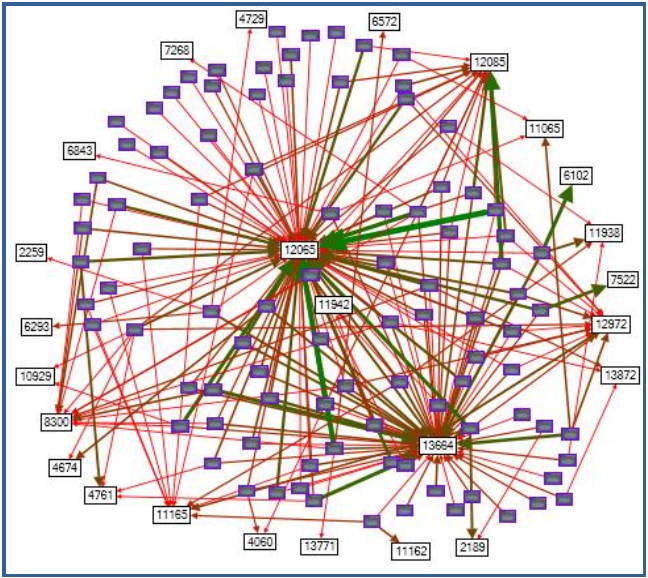

Figure 1. Directed network graph for sample Jazz project showing highly dense network segments for practitioners "12065" and "13664".

*2) Procedures and Metrics*

Software tasks were planned in multiple iterations for each project area (P1 – P10 in Table I). However, the number of iterations varied for each project (e.g., P3 tasks were completed in two iterations, whereas P5 tasks were executed in 17 iterations). To examine whether practitioners' behaviors changed over project duration we therefore divided each project's tasks and artifacts into four quarters (start, early-mid, late-mid, and end). Teams' attitudes and behaviors were studied using linguistic analysis (see subsection C). The various project areas in P1 to P10 were used to uniquely identify the nature of the software tasks; e.g., those working on P1 and P2 in Table 1 were developing UI components and undertaking usability-related tasks. We used project area and task type as our units of analysis, and also drew comparisons over project phases and at the Jazz organization level.

## C. Lingisitic Analysis Techniques

Previous research has identified that individual linguistic style is quite stable over time and that text analysis programs are able to accurately link language characteristics to attitudes and behavioral traits (see [28], for example). We employed the Linguistic Inquiry and Word Count (LIWC) software tool in our analysis. The LIWC is a software tool created after four decades of research using data collected across the USA, Canada and New Zealand [29]. This tool captures over 86% of the words used during conversations (around 4500 words). Written text is submitted as input to the tool in a file that is then processed and summarized based on the LIWC tool's dictionary. Each word in the file is searched for in the dictionary, and specific type counts are incremented based on the associated word category after which summary output is provided. These different dimensions in the summary are said to capture the psychology of individuals by assessing the words they use [28-29]. We provide a summary of the LIWC linguistic categories that were considered, along with brief theoretical justifications for their inclusion, in Table II.

TABLE I. SUMMARY STATISTICS FOR THE SELECTED JAZZ PROJECTS

| Project ID | WI Count | Project Area (nature of task) | Total Contributors | Total Messages | Period (days) |
|---|---|---|---|---|---|
| P1 | 54 | User Experience – tasks related to UI development | 33 | 460 | 304 |
| P2 | 112 | User Experience – tasks related to UI development | 47 | 975 | 630 |
| P3 | 30 | Documentation – tasks related to Web portal documentation | 29 | 158 | 59 |
| P4 | 214 | Code (Functionality) – tasks related to application development | 39 | 883 | 539 |
| P5 | 122 | Code (Functionality) – tasks related to application development | 48 | 539 | 1014 |
| P6 | 111 | Code (Functionality) – tasks related to development of application middleware | 25 | 553 | 224 |
| P7 | 91 | Code (Functionality) – tasks related to development of application middleware | 16 | 489 | 360 |
| P8 | 210 | Project Management – tasks under the project managers' control | 90 | 612 | 660 |
| P9 | 50 | Code (Functionality) – tasks related to application development | 19 | 254 | 390 |
| P10 | 207 | Code (Functionality) – tasks related to development of application middleware | 48 | 640 | 520 |
| ∑ | 1201 | - | 394 | 5563 | - |

For example, consider the following sample comment:

"We are aiming to have all the patches ready by the end of this release; this will provide us some space for the next one. Also, we are extremely confident that similar bug-issues will not appear in the future."

In this comment the author is expressing optimism that the team will succeed, and in the process finish ahead of time and with acceptable quality standards. In this quotation, the words "we" and "us" are indicators of team or collective focus, "all", "extremely" and "confident" are associated with certainty, while the words "some" and "appear" are indicators of tentative processes. Words such as "bug-issues" and "patches" are not included in the LIWC dictionary and would not affect the context of its use – whether it was to indicate a fault in software code or a problem with one's immunity to, and treatment for, a disease. Although these omissions may be thought to represent a confounding factor, we know that the context is software development; and what *is* of interest, and *is* being captured by the tool, is evidence of attitudes and behaviors. As noted in Section II, previous work has provided confirmation of the utility of the LIWC tool for examining behaviors [20], and we have also uncovered insightful findings in this study. In this work we examine whether practitioners' behaviors change over project duration and how certain tasks affect teams' attitudes and behaviors along multiple linguistic dimensions.

TABLE II. SUMMARY OF THE LINGUISTIC DIMENSIONS

| Linguistic Category | Abbreviation (Abbrev.) | Examples | Reason for Inclusion |
|---|---|---|---|
| Pronouns | I | I, me, mine, my, I'll, I've, myself, I'm | Elevated use of first person plural pronouns (we) is evident during shared situations, whereas, relatively high use of self references (I) has been linked to individualistic attitudes [30]. Use of the second person pronoun (you) may signal the degree to which members rely on (or delegate to) other team members [31]. |
| | we | we, us, our, we've, lets, we'd, we're, we'll | |
| | you | you, your, you'll, you've, y'all, you'd, yours, you're | |
| Cognitive language | insight | think, consider, determined, idea | Software teams were previously found to be most successful when many group members were highly cognitive and were natural solution providers [32]. These traits are also linked to effective task analysis and brainstorming capabilities. |
| | discrep | should, prefer, needed, regardless | |
| | tentat | maybe, perhaps, chance, hopeful | |
| | certain | definitely, always, extremely, certain | |
| Work and Achievement related language | work | feedback, goal, boss, overtime, program, delegate, duty, meeting | Individuals most concerned with task completion and achievement are said to reflect these traits during their communication. Such individuals are most concerned with task success, contributing and initiating ideas and knowledge towards task completion [10]. |
| | achieve | accomplish, attain, closure, resolve, obtain, finalize, fulfill, overcome, solve | |
| Leisure, social and positive language | leisure | club, movie, entertain, gym, party, jog, film | Individuals that are personal and social in nature are said to communicate positive emotion and social words and this trait is said to contribute towards an optimistic group climate [10]. Leisure related language may also be an indicator of a team-friendly atmosphere. |
| | social | give, buddy, love, explain, friend | |
| | posemo | beautiful, relax, perfect, glad, proud | |
| Negative language | negemo | afraid, bitch, hate, suck, dislike, shock, sorry, stupid, terrified | Negative emotion may affect team cohesiveness and group climate. This form of language shows discontent and resentment [33]. |

## 4. RESULTS AND ANALYSIS

### A. Changes Over Project Duration

#### 1) Messages - Communication Patterns

First, we consider the way teams' messages are distributed over the four project phases: start, early-mid, late-mid, and end. Fig. 2 shows that practitioners generally communicated more frequently on project tasks during the beginning and end phases of their projects. These practitioners tended to exchange a more stable and consistent mean number of messages per WI in the early-mid and late-mid stages of their projects.

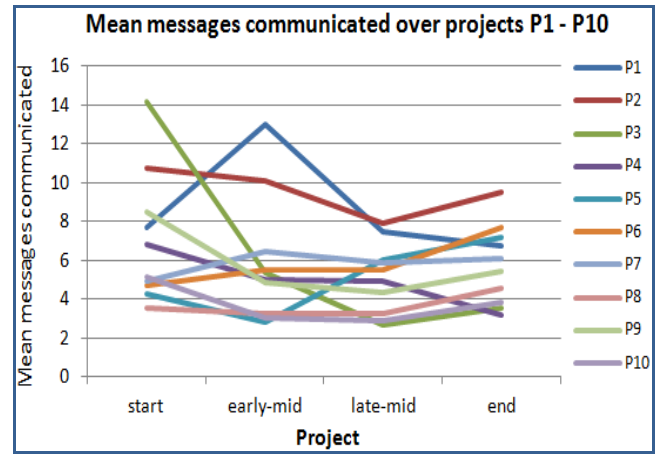

Figure 2. Mean messages communicated over project phases.

#### 2) Linguistic Analysis – Behavior Patterns

Second, we report our results for some of the linguistic dimensions in Table II; due to space limitations we provide visualizations across the four project phases for five linguistic dimensions in seven of the ten projects (representing projects that belong to the four types – user experience, documentation, coding, and project management) in Fig. 3 to depict team behaviors spanning the duration of software projects.

Overall, Fig. 3 shows that while practitioners' use of self references (e.g., I, me, my) fluctuated over their projects, incidence of this type of language generally reduced towards project completion. Although not included in Fig. 3, we also observe that measures for collective (e.g., we, our, us) and reliance (e.g., you, your, you're) language were slightly different, tending to be lower overall. Additionally, while there was more use of delegation type language at the start of the coding and project management projects (P5, P7, P8 and P9), the opposite was evident for the user experience and documentation projects (P1, P2 and P3). These patterns were then somewhat reversed for collective language use during these same projects.

We further note that practitioners were most insightful (using, for example, "think", "believe", "consider") at the start of their projects, and there was also greater discrepancy (e.g., should, would, could) in the projects' initial stages (see

Fig. 3 for illustration). Discrepancy was particularly pronounced during the early project phases for those working on coding-intensive tasks. Fig. 3 illustrates that while practitioners working on user experience related tasks became less tentative (using "maybe", "perhaps", "apparently") as their projects progressed, use of this form of language fluctuated for those working on the other tasks, and was particularly high at the start and end for those working on documentation projects. There was less variation over the different project phases for certainty type language (e.g., definitely, extremely, always) which remained low when compared to the other cognitive dimensions. Fig. 3 shows that practitioners working on the project management tasks were consistently focused on work (e.g., feedback, goal, delegate) and achievement (e.g., accomplish, attain, resolve), whereas those working on user experience, coding and documentation tasks were more focused on work and achievement in the middle and end phases of their projects. There was no consistent linguistic pattern evident for the leisure dimension (e.g., club, movie, party) during these projects, which, although low overall, tended to fluctuate over the projects.

### B. Teams' Attitudes and Behaviors

Given our intent to unearth variances in attitude and behavior among those undertaking different forms of software tasks, we separated the four types of tasks in Table I – user experience (UE), documentation (Doc), coding (Code) and project management (PM) – and conducted Kolmogorov-Smirnov normality tests for the 13 linguistic dimensions shown in Table II. Results from these tests revealed that our data violated the normality assumption. We therefore carried out Kruskal-Wallis tests to check for differences in the 13 linguistic dimensions between those undertaking the four forms of software tasks (UE, Doc, Code and PM). Table III provides these results which shows that there were significant differences ($p = 0.000$) in language usage for each of the 13 linguistic dimensions among those working in the four project areas (UE, Doc, Code and PM). We therefore conducted paired comparisons using the Mann-Whitney U test to reveal differences between type pairs (see the results in Table IV).

Considering the four types of software tasks, Table III shows that there was higher individualistic language use (e.g., I, me, my) among those working on coding intensive tasks. We observed significant differences in Table IV for this form of language use when we made comparisons with those working on user experience and project management related tasks ($p = 0.000$ and $p = 0.000$ respectively); a similar pattern was seen for those working on documentation tasks. Table III shows that those working on user experience related tasks used the lowest amount of collective language (e.g., we, our, us), and Table IV reveals that these differences were particularly pronounced when compared to those working on coding and project management related tasks ($p = 0.029$ and $p = 0.000$ respectively). Table III also shows that reliance language (e.g., you, your, you're) was highest when members were working on documentation tasks, and lowest among those working on project management tasks. Our Mann-Whitney test results revealed statistically significant differences for this dimension between those working on documentation tasks and those working on user experience, project management and coding tasks ($p = 0.004$, $p = 0.000$ and $p = 0.008$ respectively).

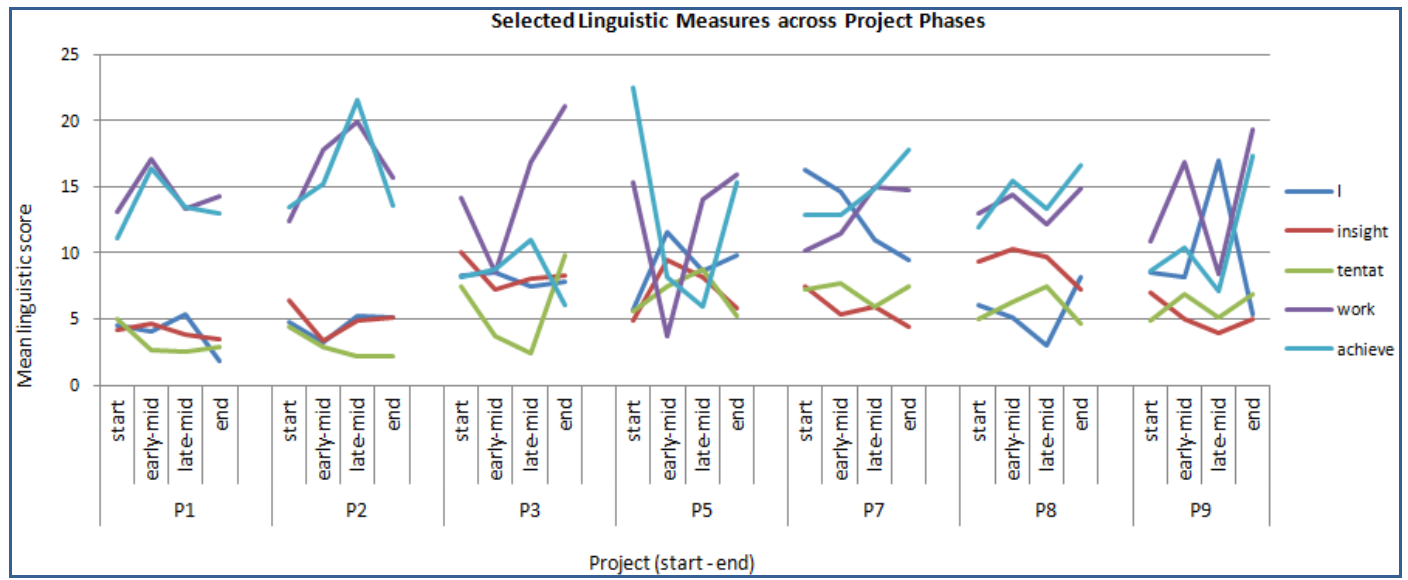

Figure 3. Linguistic measures across project phases for projects P1, P2, P3, P5, P7, P8 and P9.

Table III shows that those working on documentation tasks used the highest level of insightful language (e.g., think, believe, consider); the Mann-Whitney tests results in Table IV revealed significant differences for paired comparisons between these members and those of the user experience, coding and project management groups ($p = 0.000$, $p = 0.000$ and $p = 0.003$ respectively). On the other hand, Tables III and IV show that individuals involved with coding intensive tasks used more discrepancy (e.g., should, would, could) ($p = 0.000$ and $p = 0.000$) and tentative (e.g., maybe, perhaps, apparently) language (p = 0.000 and p = 0.001) than those working on user experience and project management tasks, respectively. Tables III and IV show that measures for certainty type words (e.g., definitely, extremely, always) were significantly higher for coders and project managers than user experience members (p = 0.000 and p = 0.000 respectively). Contributors on user experience tasks were least work (e.g., feedback, goal, delegate) and achievement (e.g., accomplish, attain, resolve) focused. Tables III and IV reveal that leisure related language use (e.g., club, movie, party) was significantly higher among project management members when compared to those working on user experience (p = 0.000) and coding (p = 0.000) related tasks.

Social language use (e.g., give, buddy, love) was particularly low among those working on user experience tasks when compared to those of the documentation (p = 0.000), coding (p = 0.000) and project management (p = 0.000) groups. On the other hand, those working on user experience tasks utilized significantly more positive (e.g., beautiful, relax, perfect) language than those working on the documentation, coding and project management tasks (p = 0.000, p = 0.000 and p = 0.000 respectively). Finally, Tables III and IV also reveal that those working on coding tasks used significantly higher amounts of negative language (e.g., afraid, hate, dislike) when compared to user experience (p = 0.000) and project management (p = 0.000) members.

TABLE III.  MEAN RANKS AND KRUSKAL-WALLIS TEST RESULTS

| Linguistic Category | Abbrev. | Mean Rank | | | | Kruskal-Wallis Test (*p*-value) |
| --- | --- | --- | --- | --- | --- | --- |
| | | UE | Doc | Code | PM | |
| Pronouns | I | 2512.210 | 2946.610 | 2928.010 | 2622.210 | 0.000 |
| | we | 2706.390 | 2815.310 | 2786.560 | 2980.560 | 0.000 |
| | you | 2780.780 | 3061.150 | 2808.000 | 2623.980 | 0.000 |
| Cognitive | insight | 2629.930 | 3267.150 | 2815.310 | 2884.120 | 0.000 |
| | discrep | 2448.830 | 2743.030 | 2954.110 | 2679.760 | 0.000 |
| | tentat | 2453.850 | 2963.810 | 2932.120 | 2732.170 | 0.000 |
| | certain | 2625.400 | 2786.210 | 2852.720 | 2813.260 | 0.000 |
| Work and Achievement | work | 2627.660 | 2914.140 | 2819.320 | 2959.130 | 0.000 |
| | achieve | 2600.950 | 2601.930 | 2851.570 | 2924.690 | 0.000 |
| Leisure, social and positive | leisure | 2730.610 | 2930.270 | 2764.970 | 3013.040 | 0.000 |
| | social | 2568.760 | 3131.470 | 2849.240 | 2876.220 | 0.000 |
| | posemo | 3391.600 | 2654.940 | 2591.400 | 2489.460 | 0.000 |
| Negative | negemo | 2568.590 | 2768.430 | 2889.090 | 2750.640 | 0.000 |

TABLE IV.  MANN-WHITNEY TEST RESULTS

| Linguistic Category | Abbrev. | Mann-Whitney Test (*p*-value) | | | | | |
| --- | --- | --- | --- | --- | --- | --- | --- |
| | | UE - Doc | UE - Code | UE - PM | Doc - Code | Doc - PM | Code - PM |
| Pronouns | I | **0.000** | **0.000** | 0.068 | 0.868 | **0.009** | **0.000** |
| | we | 0.250 | **0.029** | **0.000** | 0.768 | 0.149 | **0.000** |
| | you | **0.004** | 0.454 | **0.003** | **0.008** | **0.000** | **0.000** |
| Cognitive | insight | **0.000** | **0.000** | **0.000** | **0.000** | **0.003** | 0.271 |
| | discrep | **0.002** | **0.000** | **0.000** | 0.055 | 0.560 | **0.000** |
| | tentat | **0.000** | **0.000** | **0.000** | 0.792 | 0.058 | **0.001** |
| | certain | 0.063 | **0.000** | **0.000** | 0.491 | 0.793 | 0.454 |
| Work and Achievement | work | **0.020** | **0.000** | **0.000** | 0.429 | 0.756 | **0.032** |
| | achieve | 0.763 | **0.000** | **0.000** | **0.034** | **0.010** | 0.264 |
| Leisure, social and positive | leisure | **0.036** | 0.315 | **0.000** | 0.070 | 0.472 | **0.000** |
| | social | **0.000** | **0.000** | **0.000** | **0.022** | 0.082 | 0.645 |
| | posemo | **0.000** | **0.000** | **0.000** | 0.556 | 0.169 | 0.107 |
| Negative | negemo | **0.017** | **0.000** | **0.000** | 0.228 | 0.878 | **0.010** |

## 5. DISCUSSION AND IMPLICATIONS

### A. How do globally distributed agile software practitioners' behaviors change over project duration?

Jazz team members engage each other most frequently at project start and project completion. This finding is not particularly surprising given the need to establish overall goals and work assignments at the beginning of a project and then to intensively assess the project at closure to ensure that the features developed match those requested. More specifically, we found that Jazz teams became more collective as their projects progressed, an indicator that these teams were operating cohesively in the norming and performing phases of group work [34]. Teams tend to evolve collectively after overcoming initial differences and conflicts, and elevated levels of collective behaviors is an indicator of more shared and established team norms.

In line with the higher volume of messages exchanged, we also found the highest levels of cognitive language use at the start of these Jazz projects, when project initiation and scoping were conducted. Higher levels of cognitive processes have been linked previously to higher task performance [32]. Practitioners' intensive involvement in their teams' knowledge network at this time may be beneficial for ensuring their individual contextual awareness during the remaining stages of the project. Additionally, the less knowledgeable members may benefit the most from these early exchanges, when the projects' top members are available and can readily engage and share their knowledge with the wider team. The higher levels of engagement towards project completion (which perhaps comprise urgent and frequent reminders of work remaining and reflections) may also aid the less knowledgeable and less aware members regarding their future involvement, and particularly, those features that may be potentially reusable. There was consistent work and achievement language use among those involved with project management tasks. Work and achievement focus is said to be a necessary ingredient for agile practitioners, and particularly for those that need to perform across roles and self-organize [35]. Project managers can encourage this approach to work among their wider team members, which might explain the evidence for consistent work and achievement focus among those operating on the project management tasks.

We observe that coders and those involved with project management tasks relied on each other the most in the early project stages, while those working in other areas became more inter-dependent as their projects progressed. We expected that agile distributed teams employing an iterative approach (such as those working on/with Jazz) would establish and firm up project requirements, task priorities and task assignments in the early stages and then continually as projects progress. In particular, for documentation and user experience tasks, we anticipated that these teams would become most active after the more computationally intensive features are delivered by the coders. This was evident for the ten teams studied here. There seems to be more stability in terms of task assignment at the start of the project for coders and those working around project management tasks, while the others tend to be stable after early features are delivered and the project is properly established. This finding has implications for human resource management strategies (discussed below).

## B. How does the nature of software tasks affect globally distributed agile teams' attitudes and behaviors?

Our results show that software team members' attitudes and behaviors are influenced by the nature of the tasks they perform, indicating that recommendations from role theories [10] and psychology [24] are indeed relevant to software engineering and may augment human resource strategies regarding task assignments. Software practitioners were found to express particular attitudes when working on certain forms of software features. We also observed consistent patterns across project areas for some of the linguistic dimensions, suggesting that Jazz software practitioners express some distinct attitudes and behaviors regardless of the localized nature of the work they are performing. As a group, the IBM Rational Jazz developers were more self-focused than they were collective, and these practitioners did not rely excessively on each other. While they expressed limited certainty, these practitioners were highly task and achievement focused regardless of the task they were undertaking. Additionally, although these individuals were task focused, they also communicated with high amounts of positive and social language and expressed little frustration during their discourses. We now examine in detail the specific behaviors evident among the four groups of practitioners.

### 1) Attitudes Among Coders

Jazz practitioners working on coding-intensive tasks were highly self-focused and, when compared to those addressing other software features, coders exhibited the highest use of cognitive processes. Less desirable negative emotion was also most common among coders. Although there was evidence of greater individual ownership when practitioners were working on coding-intensive or computational tasks, members working on such tasks were more passionate during their engagement. We observed that coders offered more suggestions to others, but they were also conservative at times. At other times these individuals were highly confident, but equally frustrated. We believe that the more pronounced use of several language dimensions by Jazz practitioners involved in coding intensive tasks may be indicative of the cognitive and mental challenges involved in such activities [36], and has implications for teams' task assignments. Individuals involved in coding may be required to possess higher levels of cognitive skills than those working on other tasks. Given our findings, we suspect that individuals who are eager and able to work independently may also be likely to favor such tasks.

### 2) Attitudes Among Documenters

We found Jazz practitioners working on documentation tasks to be self-focused, but these individuals also delegated and relied heavily on each other, and demonstrated high levels of group or collective focus. In addition, those dealing with documentation were insightful, but showed tentativeness at times. Those working on documentation tasks maintained work focus, and were social and positive (somewhat aligned with the group or people focus noted earlier), tending to engage in more off-task discussions than other members. Based on these observations we contend that those involved in documentation activities may need less help, but they may also need to possess higher levels of perception and be intuitive. Those that are extroverted or agreeable [33] may perform well on documentation tasks.

### 3) Attitudes Among Usability practitioners

Jazz practitioners working on user experience tasks were the least self-focused and collective of those studied, and these individuals also used the lowest levels of cognitive processes. We also observe the least work and achievement focus among usability members. However, these practitioners were extremely positive, tending to show very little frustration, confirming similar results found in our preliminary analysis of three different Jazz project areas [11]. In the current work this pattern is maintained across multiple usability related projects. Those in Jazz that were working on usability projects were constructive, tending to be much more optimistic and upbeat and less negative. Thus, practitioners that are emotionally stable may perform well on such tasks [33], particularly in a distributed software team where there is a need to solicit feedback about software features' 'look and feel' from unfamiliar team members.

*4) Attitudes Among Project Managers*

It is commonly held that project managers favoring an extroverted, confident and inquisitive outlook are most successful [32]. We found evidence in support of this position, as we noted the highest use of collective processes among this group of practitioners. These individuals were also highly cognitive whenever they communicated and were achievement driven. We examined the role distribution of teams and found twice the number of project managers and team leaders working among those undertaking project management tasks compared to those working on the other forms of task. In Jazz, those working on project management tasks exhibited a team outlook, and were also intuitive and communicated with confidence. These traits may be helpful for these individuals, and may promote general staff (followers') confidence.

We anticipated that our findings may be linked to the individuals assigned to these teams, as against the nature of the problems or features they addressed, such that the differences in linguistic patterns observed may be due to the differences in team membership. We therefore checked for similarities in team membership for those performing the various software tasks. We found similarities in practitioners' participation across the various software tasks. While 62% and 81% of those working on the user experience tasks were also involved in the project management and coding tasks, respectively, 52%, 69% and 93% of those working on documentation tasks were also working on user experience, project management and coding tasks, respectively. Finally, 63% of those working on the project management tasks were also involved in coding tasks. These overlaps in the membership of teams performing these tasks support the statistical findings above that team behaviors vary given their project environment.

## C. Implications for Project Governance and Tool Design

Agile global practitioners need to possess and draw on various skills when they are working in a high performing team such as the Jazz teams studied here (tools created by Jazz teams are being used (and were positively reviewed) by over 30,000 companies – see jazz.net). These requirements may be most critical for distributed teams where the opportunity to engage face-to-face (F-t-F) is not available. In line with the early assessment of DeMarco and Boehm [23], agile practitioners need to possess high levels of intra-personal, organizational and inter-personal skills if they are to succeed. Those solving usability related tasks may need to demonstrate high levels of inter-personal and social skills, while coders need to show higher level of organizational and intra-personal skills. Project managers or those engaged in the project management team need inter-personal, organizational and intra-personal competence to facilitate teams' success.

These requirements have implications for staff placement, and assigning the right individuals to tasks. One solution is for project managers to select highly skilled practitioners possessing a mix of all the necessary skills (inter-personal, organizational and intra-personal) regardless of the software tasks being undertaken. A second solution is to assign a mix of those with specific specialties (inter-personal or organizational or intra-personal) to each team. Each of these strategies will have implications for project cost and team function. In the first instance project managers may be faced with higher human resource budgets, as those with multiple talents will likely demand higher remuneration. Moreover, the availability of across-the-board high performers is inherently limited. On the other hand, the second scenario may result in underutilization of human resource, as there would be times when there is less need for specific expertise (as was seen for Jazz teams working on user experience and documentation tasks in this study). Project managers will need to make definitive decisions, as using a 'one size fits all' model may not be suitable in agile distributed contexts.

Additionally, the higher levels of engagement at project initiation and completion have implications for global agile team members' availability and their learning. If team members are not available at these critical times they stand to lose contextual awareness information – both knowledge from project initiation and scoping, and reflections and evaluations aimed at assessing that features match requirements at project completion. Information shared at these times will comprise teams' ongoing tacit knowledge that may not be evident or captured in other forms. Project managers should ensure team member participation during these key project phases, since this is the time when members are likely to benefit the most from their teams' knowledge pools. This has particular significance for distributed software teams, where distance (both geographical and temporal) may affect software practitioners' availability at these critical moments.

Finally, the inclusion of psycholinguistic analysis and reporting functionality in Jazz (or similar tools) should help project managers to monitor team members' attitudes and behaviors. Managing this information, like any other process data (e.g., task assignment and effort), should help a project manager using Jazz to take appropriate action around task assignment and team composition, and avoid conflict and performance issues. In order to be reliable and accurate, such a tool will need to incorporate data mining principles (particularly data pre-processing techniques) and tested natural language processing methods. For instance, in monitoring the user experience project area for the Jazz teams studied here, a very high level of negative language use (e.g., afraid, hate, dislike) not typical of those working on usability features would be indicative of poor team synergies. If left unchecked, this may lead to conflict and staff turnover. Thus, awareness of teams' frustration supported by the provision of behavioral or mood visualizations would allow the project manager to assign new and more flexible members with high levels of inter-personal skills to these tasks. Similarly, if those working on the project management team are highly tentative, this may be a sign to call a project meeting to promote their confidence, as uncertain project governance may also lead to inadequate project performance.

## 6. LIMITATIONS

*Measurement Validity*: The LIWC language constructs used to measure teams' attitudes and behaviors have been used previously to study this subject, and were assessed for validity and reliability [28]. However, although the LIWC dictionary was able to capture 66% of the overall words used by Jazz practitioners, the adequacy of these constructs in the specific context of software development warrants further investigation. Nonetheless, we checked a small sample of the messages to see what might account for the remaining words being ignored by the LIWC tool and found that there were large amounts of highly specialized software development related language (e.g., J2EE, LDAP, JACC, API, XML, TAME, JASS, Jazz, URI, REST, HTTP) during Jazz practitioners exchanges. Moreover, what was of interest, and was captured by the LIWC tool, was evidence of attitude, demeanor and behavior. Finally, communication was assessed based on messages sent around software tasks. These messages were extracted from Jazz, and may not represent all of the project teams' communications. Offsetting this concern is the fact that, as Jazz was developed as a globally distributed project, developers were required to use messages so that all other contributors (irrespective of their physical location) were aware of product and process decisions regarding each WI.

*Internal and External Validity*: Although we achieved data saturation after analyzing the third project case, the tasks, history logs and messages from the ten projects (out of 94) may not necessarily represent all the teams' processes in the repository. Additionally, work processes and work culture at IBM are also specific to that organization and may not be representative of organization dynamics elsewhere.

## 7. CONCLUSIONS AND FUTURE WORK

Recent evidence highlights ongoing problems in software project performance, which is said to be linked to communication and behavioral issues. Accordingly, there is growing emphasis on studying human-related processes to provide recommendations for software process improvements. This drive to understand the human aspects of software development is particularly relevant for agile and global endeavors given these approaches emphasize the people dimension. We have followed this line of research and have employed psycholinguistic analysis to study the way IBM Rational Jazz globally distributed agile practitioners' behaviors change over project duration and how software teams' attitudes and behaviors vary given the tasks they were undertaking.

We found the highest levels of project engagement at project start and completion, and increasing levels of team collectiveness as Jazz projects progressed. Additionally, we found that Jazz practitioners expressed the most ideas at the time of project scoping. However, overall, Jazz teams' attitudes and behaviors varied given the nature of the tasks they were performing. Our results have implications for software project governance, and these findings may also inform new requirements for extending IBM Rational Jazz or similar tools.

Our next step is to examine the way specific project members (top members and project leaders) contribute to the dynamics of these teams. In the short term, we also plan to consider group dynamics at a more granular task level (e.g., for bugs, new features and feature enhancements), and to complement our linguistic analysis with more contextual bottom-up analysis techniques. Our longer term goal is to provide tool support for teams' behavior visualizations.

## ACKNOWLEDGMENTS

We thank IBM for granting us access to the Jazz repository. S. Licorish is supported by an AUT Vice-Chancellor's Doctoral Scholarship Award.